# Molybdenum diselenide-manganese porphyrin bifunctional electrocatalyst for hydrogen evolution reaction and selective hydrogen peroxide production

Antonia Kagkoura,[1*] Christina Stangel,[1] Raul Arenal,[2,3,4*] Nikos Tagmatarchis[1*]

[1] Theoretical and Physical Chemistry Institute, National Hellenic Research Foundation, 48 Vassileos Constantinou Avenue, 11635 Athens, Greece

[2] Laboratorio de Microscopias Avanzadas (LMA), Universidad de Zaragoza, Mariano Esquillor s/n, 50018 Zaragoza, Spain

[3] Instituto de Nanociencia y Materiales de Aragon (INMA), CSIC-U. de Zaragoza, Calle Pedro Cerbuna 12, 50009 Zaragoza, Spain

[4] ARAID Foundation, 50018 Zaragoza, Spain

ABSTRACT

Electrochemical reactions for hydrogen and hydrogen peroxide production are essential for energy conversion to diminish energy crisis, but still lack efficient electrocatalysts. Development of non-noble metal bifunctional electrocatalysts for hydrogen evolution and $2e^-$ oxygen reduction reaction to ease reaction kinetics is a challenging task. Integration of single components by employing easy strategies provides a key-step towards the realization of highly active electrocatalysts. In this vein, $MoSe_2$ owns catalytic active sites and high specific surface area but suffers from insufficient conductivity and high catalytic performance that noble-metals provide. Herein, $MoSe_2$ was used as a platform for the incorporation of manganese-metallated porphyrin (MnP). The developed hybrid, namely $MoSe_2$-MnP, by the initial metal-ligand coordination and the subsequent grafting with MnP was fully characterized and electrochemically assessed. The bifunctional electrocatalyst lowered the overpotential toward hydrogen evolution, improved reaction kinetics and charge transfer processes and was extremely stable after 10,000 ongoing cycles. Simultaneously, rotating ring disk electrode analysis showed that oxygen reduction proceeds through the $2e^-$ pathway for the selective production of hydrogen peroxide with a high yield of 97%. The new facile modification route can be applied in diverse transition metal dichalcogenides and will help the development of new advanced functional materials.

**Introduction**

Water electrolysis is a promising tool for carbon-free hydrogen ($H_2$) production towards the commercialization of hydrogen economy. However, lack of low-cost electrocatalysts for alkaline hydrogen evolution reaction (HER) to replace scarce and pricy Pt-based ones prevents widespread usage of water splitting. At the same time, hydrogen peroxide ($H_2O_2$) can have double role in fuel cell technology as alternative liquid oxidant in place of gaseous oxygen and as a fuel.[1,2] Additionally, $H_2O_2$ is a valuable oxidative chemical with rapidly increasing demand in applications, including chemical synthesis, pulp/paper bleaching, medical and textile industry and wastewater treatment.[3] Molecular oxygen can be electrochemically reduced to water via a $4e^-$ pathway, or hydrogen peroxide with $2e^-$ transferred in aqueous media. Moreover, selective $2e^-$ oxygen reduction reaction (ORR) provides a green and mild pathway to produce $H_2O_2$ compared to the traditional energy, capital, and waste intensive anthraquinone process. Both HER and $2e^-$ ORR need novel non-noble metal electrocatalysts to speed up the reactions to support large-scale supporting infrastructures for electrochemical hydrogen and hydrogen peroxide production.

Transition metal dichalcogenides (TMDs) are promising alternatives to Pt due to their inherent properties stemming from the exposed prismatic edges, abundance and low cost.[4] Molybdenum diselenide ($MoSe_2$) is less studied than molybdenum disulfide ($MoS_2$) and has exhibited relatively superior electrocatalytic activity and chemical stability.[5-7] It is well known that the 2H-phase edge sites are mostly catalytically active for HER, opposed to 1T-phase, where unsaturated Mo atoms at the basal plane of $1T-MoS_2$ are electrocatalytically active.[8] Additionally, bottom up approaches based on hydrothermal/solvothermal reaction of metal and chalcogen molecular precursors lead to high yield production of TMDs of both polymorphs showing plenty exposed edges and increased HER electrocatalytic activity.[7,9,10] However, combination of $MoSe_2$ and generally of TMDs with other species and/or substrates is needed in order to facilitate charge transfer, improve conductivity and overall electrocatalytic performance.[13-17] In the meantime, porphyrins are large $\pi$-conjugated nitrogen-containing aromatic macrocycles with a high degree of electronic delocalization. Actually, hybridization of $MoSe_2$ with a free porphyrin led to improved electrocatalytic HER performance.[18] Additionally, porphyrins provide an efficient and low-cost metal carrier with auspicious electrocatalytic activity for both HER and ORR.[19,20] On the other hand, TMDs show rather low performance for electrocatalytic ORR, therefore association with other species or atom doping is mandatory for efficient electrocatalysis.[21-23] Most importantly, the realization of effective bifunctional electrocatalysts through robust and easy modification approaches for TMDs is of high importance.

Despite their interesting qualities, functionalization of TMDs is mandatory to fully exploit their properties and prepare advanced functional materials. Towards this road, various modification routes have been employed to modify both phases involving both covalent or non-covalent approaches.[24,25] Edge-located sulfur vacancies were exploited for the covalent functionalization of both polymorphs.[26,27] Notably, for chemically exfoliated 1T-MoS$_2$, material functionalization mostly rests on electron-rich sulfur species at the basal plane,[28,29] while for 2H-MoS$_2$ an external supply of electrons is needed.[30] However, neutralization of negative charges[31] can have negative effect in electrocatalysis. Therefore, versatile and effective modification routes that can be applied to both phases, without disrupting the electronic network of TMDs, still need to develop towards the realization of functional materials for electrocatalysis.

To the best of our knowledge there is no report for the combination of MoSe$_2$ with metalloporphyrins. In this work, we initially employed an easy functionalization route for the modification of MoSe$_2$ with aminopyridine, through coordination, for the subsequent covalent grafting with carboxy terminated manganese(III) porphyrin (MnP), to obtain MoSe$_2$-MnP. The MoSe$_2$-MnP hybrid material was tested against HER and 2e$^-$ ORR in alkaline media and has proven to be very efficient and stable bifunctional electrocatalyst. In fact, MoSe$_2$-MnP showed lower HER overpotential of 230 mV compared to intact MoSe$_2$, excellent stability after 10,000 ongoing cycles and small charge transfer resistance. At the same time, $MoSe_2$-MnP exhibited high selectivity of 97% toward hydrogen peroxide production and negligible current loss after 10,000 seconds. The current work goes beyond the state of the art, as it uses a new functionalization route that can be applied to various TMDs showing both polymorphs toward the development of functional nanomaterials.

**Methods**

**General.** Chemicals, reagents, and solvents were purchased from Sigma-Aldrich and used without further purification. Infrared (IR) spectra were obtained on a Fourier Transform IR spectrometer (Equinox 55 from Bruker Optics) equipped with a single reflection diamond ATR accessory (DuraSamp1IR II by SensIR Technologies). Raman measurements were recorded with a Renishaw confocal spectrometer at 514 nm. The data were obtained and analysed with Renishaw Wire and Origin software. Thermogravimetric analysis was acquired using a TGA Q500 V20.2 Build 27 instrument by TA in a nitrogen (purity >99.999%) inert atmosphere. High-resolution scanning transmission electron microscopy (STEM) imaging and energy dispersive X-ray spectroscopy (EDS) analyses have been carried out in a probe-corrected Thermo Fisher Titan-Low-Base 60-300 operating at 120 kV (equipped with a Cs-probe corrector (CESCOR from CEOS GmbH)). To filter noise, the EDS spectrum was smoothed using a Savitzky−Golay filter (second-order polynomial). For these TEM studies, the sample has been dispersed in ethanol and the suspension has been ultrasonicated and dropped onto a copper carbon holey grid. All electrochemical measurements were carried out using an Autolab PGSTAT128N potentiostat/galvanostat and were carried out at room temperature in a standard three-compartment electrochemical cell by using a graphite rod as a counter-electrode, an RDE with glassy carbon disk (geometric surface area: 0.0196 $cm^2$) as a working electrode, and Hg/HgO (0.1M KOH)) as reference electrode. LSV measurements for HER were carried out at room temperature in $N_2$-saturated aqueous 0.1M KOH The catalyst ink was prepared by dispersing 4.0 mg of the catalytic powder in a 1 mL mixture of deionized water, isopropanol, and 5% Nafion (v/v/v=4:1:0.02) and sonicated for 30 min prior use. Before casting the electrocatalytic ink on the electrode's surface, the working electrode was polished with 6, 3 and 1mm diamond pastes, rinsed with deionized water, and sonicated in double-distilled water. Afterwards, 8.5 μL aliquots of the electrocatalyst were casted on the electrode surface and were left to dry at room temperature. Electrochemical impedance spectroscopy (EIS) measurements were conducted from $10^5$ to $10^{-1}$ Hz with an AC amplitude of 0.01 V. EIS measurements for HER were conducted at a potential where significant current was recorded, corresponding to -2 mA/cm$^2$, while EIS data were fitted to Randles circuit. The EIS measurements for ORR were conducted at a potential where significant current was recorded, corresponding to -0.2 mA/cm$^2$, while EIS data were fitted to Randles circuit. The ring potential was fixed at 1.4 V vs. RHE. Based on the RRDE data, the n value and the percentage (%) of produced $H_2O_2$ can be determined by the following equations: n= 4$I_{Disk}$ / (($I_{Disk}$+ $I_{Ring}$)/N) and %$H_2O_2$ = (200 $I_{Ring}$/N)/ ($I_{Disk}$+ $I_{Ring}$/N), where $I_{disk}$ is the current of the disk electrode, $I_{Ring}$ is the current of the ring electrode, and N is the collection efficiency of the Pt ring, which was provided as 0.249 by the manufacturer. The number of electrons transferred in the reduction of one $O_2$ molecule (n) can be determined by modifying the K–L equation as follows: $1/_j = 1/_{jd} + 1/_{jk} = 1/B\omega^{1/2} + 1/_{jk}$, where ω is the angular velocity and B is K-L slope given by the following equation B = 0.20nF$C_0$$D_0^{2/3}$ν$^{-1/6}$. Here, n is the number of electrons transferred, F is the Faraday constant (F = 96485 C/mol), $D_0$ is the diffusion coefficient of $O_2$ ($D_0$ =1.9 ×$10^{-5}$ cm$^2$ s$^{-1}$), ν is the kinematic viscosity of the solution 2 (ν =0.01 cm$^2$ s$^{-1}$) and $C_0$ is the concentration of dissolved $O_2$ in the solution ($C_0 = 1.2 \times 10^{-6}$ mol $cm^3$ ). The constant of 0.2 is adopted when the rotation speed is expressed in rpm.

**MoSe₂.** In a round bottom flask, molybdenum hexacarbonyl (3 mmol) and selenium powder (5.8 mmol) were dissolved in dry p-xylene (145 mL). Afterwards the mixture was placed into a 300 mL Teflon-lined stainless-steel autoclave reactor. The autoclave reactor was kept at 250 °C for 24 h. The resulting suspension was filtrated through PTFE membrane filter (0.2 μm) and washed with acetone and dichloromethane. The solid residue was collected to obtain 792 mg of MoSe₂.

**f-MoSe₂.** MoSe₂ (30 mg) was dispersed in $H_2O$ (30 mL) and sonicated for 10 minutes. Afterwards, 4-aminopyridine (0.3375 mmol) was added in the dispersion and the dispersion was sonicated for 1 minute. The resulting suspension was filtrated through PTFE membrane filter (0.2 μm) and washed with water. The solid residue was collected to obtain 48 mg of f-MoSe₂.

**5-(4-carboxyphenyl)-10,15,20-triphenylporphyrin-Mn(III) chloride (MnP).** 5-(4-carboxyphenyl)-10,15,20-triphenyl porphyrin (15 mg, 0.02277 mmol) and $MnCl_2$ (90 mg, 0.4554 mmol) were dissolved in DMF (10 mL) and the solution was heated to reflux for 4 hours. Then, DMF was distilled under reduced pressure. The residue was dissolved in $CH_2Cl_2$ together with two drops of methanol and washed with water. The organic phase dried over $Na_2SO_4$, and concentrated to dryness. Column chromatography (SiO₂, $CH_2Cl_2$/MeOH, 90:10) and reprecipitation of the product with $CH_2Cl_2$ and cold hexane afforded the entitled MnP as greenish-brown solid (15 mg, 88%). [1]H-NMR signals were broad due to the presence of paramagnetic Mn(III). UV-Vis ($CH_2Cl_2$): *λmax* = 343, 374 ,398, 471, 526, 575, 613. ATR-IR: ṽ = 1715 (C=O carboxylic), 1608 (C=C), 1339 (C-N), 1073 (C-H), 704 (N-H) cm$^{-1}$ (porphyrin core vibrations).

**MoSe₂-MnP.** In a two-neck round bottom flask, manganese porphyrin bearing carbonyl units MnP (7.3 mg) was added in dry dichloromethane (15 mL) and sonicated for 30 min under N₂ atmosphere. Afterwards, EDCI (13.42 mg, 0.07 mmol) was added at 0 °C and the suspension was left to stir for 1 h. Then, f-MSe$_2$ (8 mg), HOBt (13.51 mg, 0.1 mmol) and N,N-diisopropylethylamine (20 μL) were added and the dispersion was left to stir for five days at room temperature. Finally, the resulting suspension was filtrated through PTFE membrane filter (0.2 μm) and washed with dichloromethane, methanol and acetone to obtain 15 mg of MoSe$_2$-MnP as powder.

**Results and discussion**

Initially, MoSe$_2$ was prepared by an easy one-pot solvothermal process employing molybdenum hexacarbonyl Mo(CO)$_6$ and Se powder as precursors.[32] Specifically, heating of Mo(CO)$_6$ and Se in p-xylene led to the growth and production of high yield MoSe$_2$ possessing metallic properties and with exposed edges. Afterwards, amino terminated groups were introduced to MoSe$_2$ via a mild and fast route, through metal-ligand coordination between Mo atoms of MoSe$_2$ and the pyridine group of *p*-aminopyridine, yielding functionalized MoSe$_2$ (f-MoSe$_2$). The upcoming condensation reaction between amino-modified MoSe$_2$ and carboxylic acid derivative of MnP led to the formation of MoSe$_2$-MnP, as described in Scheme 1.

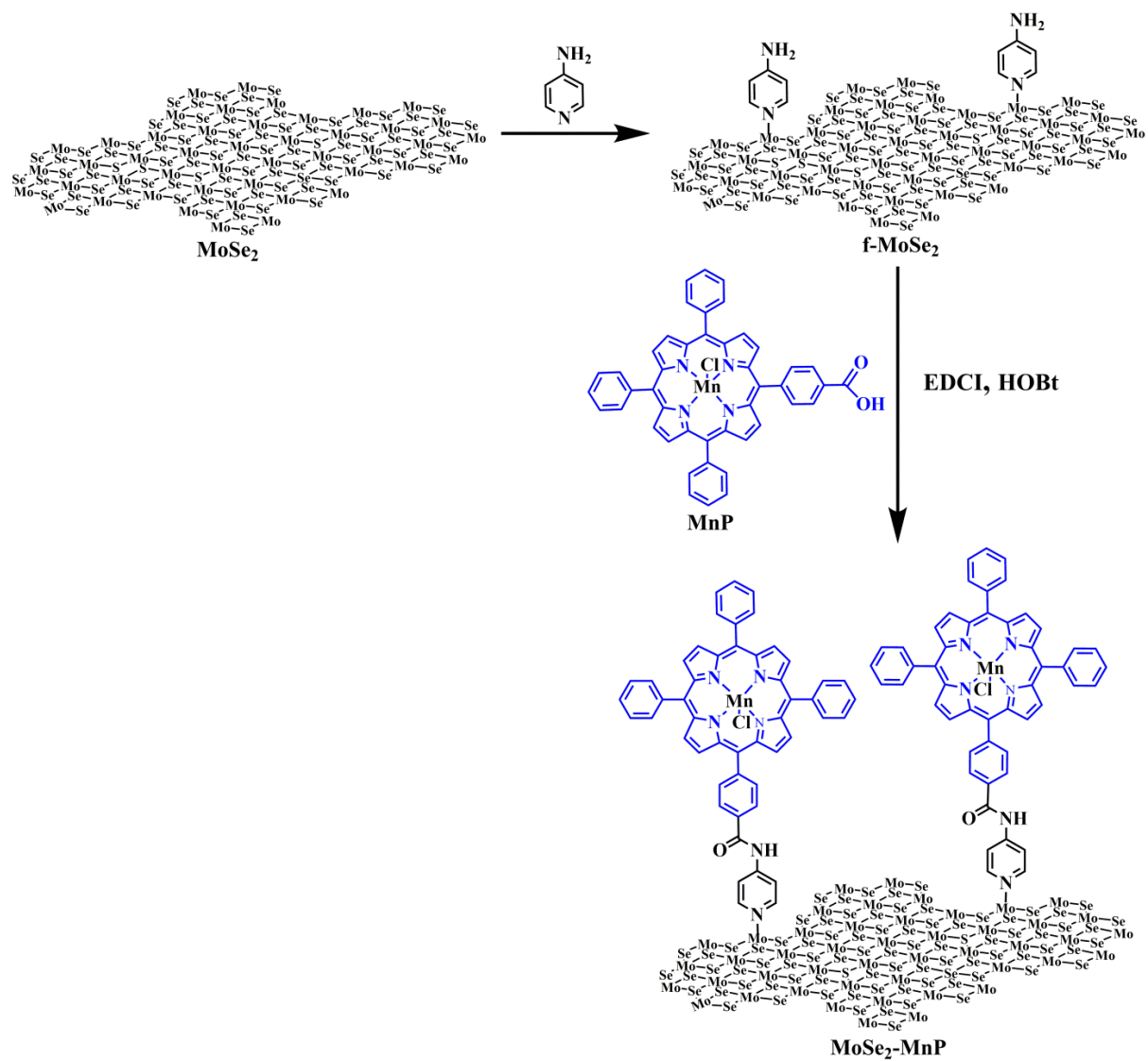

**Scheme 1.** Synthetic illustration for the functionalization of MoSe$_2$ and subsequent preparation of MoSe$_2$-MnP.

Various spectroscopic, thermal and microscopy imaging techniques were employed to follow alterations on MoSe$_2$, with end goal to verify the successful formation of MoSe$_2$-MnP. As seen in Figure 1a, the ATR-IR spectrum of f-MoSe$_2$ is governed by the characteristic viration bands due to the pyridine ring, successfully coordinated to MoSe$_2$, at 3430 (N-H), 1651 (C-N) and 1411 (C-H) cm$^{-1}$. On the other hand, the spectrum of MnP is occupied by a fingerprint carbonyl vibration mode at 1690 cm$^{-1}$ due to the -COOH units. Meanwhile, MoSe$_2$-MnP shows a discrete band at 1641 cm$^{-1}$ due to carbonyl amide stretching, deriving from the amide formed by condensation of the free amine groups being carried onto MoSe$_2$ with the carboxylic acid units of MnP (Figure 1a). Besides, N-H vibrational features are discernable in the ATR-IR spectrum of MoSe$_2$-MnP owned to the successful hybridization.

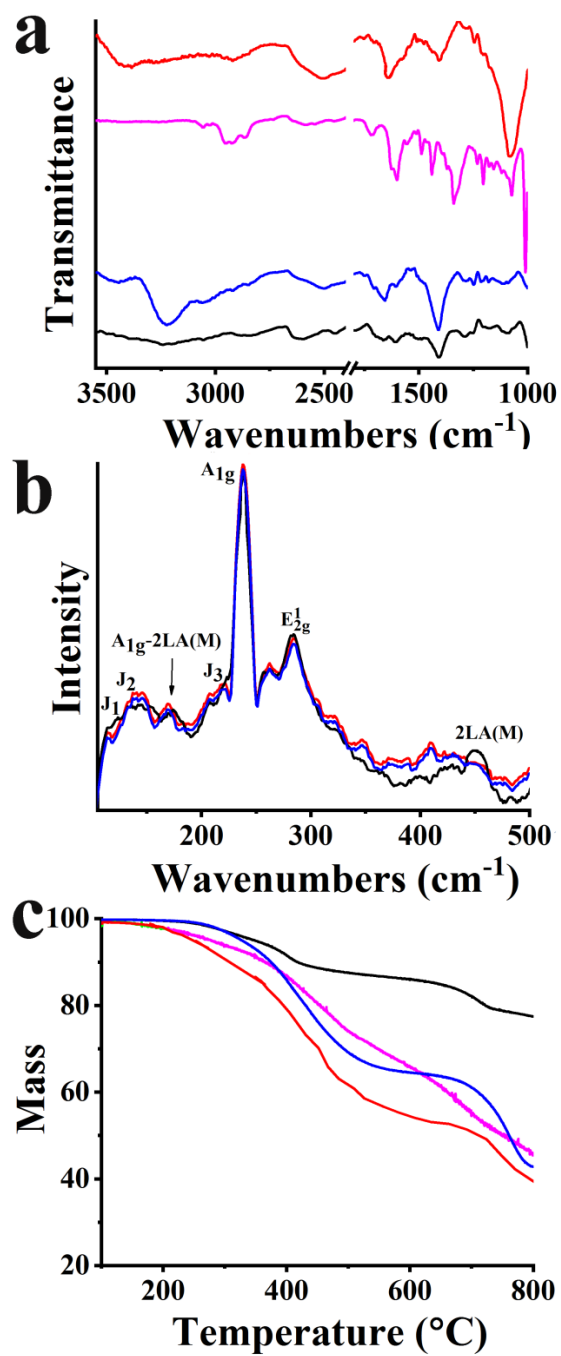

**Figure 1.** (a) ATR-IR spectra, (b) Raman spectra (514 nm), and (c) TGA graphs for MoSe$_2$-MnP, (red), MoSe$_2$ (black), f-MoSe$_2$ (blue) and MnP (pink).

Additional information on the formation of MoSe$_2$-MnP was obtained by Raman spectroscopy (Figure 1b). Specifically, the spectrum of MoSe$_2$ (excitation at 514 nm) is governed by the A$_{1g}$-

LA(M), $E^1_{2g}$, $A_{1g}$ and 2LA(M) bands, located at 171, 238, 285, 450 cm$^{-1}$, respectively. The additional $J_1$, $J_2$ and $J_3$ bands located at 118, 137 and 223 cm$^{-1}$, respectively, are fingerprint phonon modes of the metallic 1T octahedral phase of MoSe$_2$.[33,34] In the Raman spectrum of f-MoSe$_2$, the decrease of the intensity of 2LA(M), a band that is related to selenium vacancies and defect sites, compared to the one of MoSe$_2$, is associated with the successful functionalization of MoSe$_2$ as metal-ligand coordination took place at unsaturated Mo atoms. In the spectrum of MoSe$_2$-MnP there is no further alteration observed for 2LA(M), as the covalent grafting of MnP onto MoSe$_2$ has no further impact on the basal plane of the MoSe$_2$ nanosheets.

Information about the thermal stability of MoSe$_2$-MnP was obtained by thermogravimetric analysis (TGA) under N$_2$ atmosphere, while the amount of organic units covalently anchored onto MoSe$_2$ was estimated (Figure 1c). In this context, MoSe$_2$ presents a 14% mass loss up to 550 °C due to the presence of defects related to the solvothermal preparation method. On the other hand, f-MoSe$_2$ presents a much higher mass loss, up to 21%, at the same temperature region, due to the thermal decomposition of the incorporated organic species, while the thermograph of MoSe$_2$-MnP exhibits an additional mass loss of 9% due to the successful grafting of MnP onto f-MoSe$_2$. Based on the above, the loading of MnP onto the amino-modified MoSe$_2$ material was calculated to be one MnP for every 15 f-MoSe$_2$ units.

High-angle annular dark field scanning transmission electron microscopy (HAADF-STEM) imaging and energy dispersive X-ray spectroscopy (EDS-STEM) were performed for analyzing these MoSe$_2$-MnP nanostructures. Figure 2 shows the STEM studies carried out in one of these flakes. The 1T-MoSe$_2$ structures are clearly observed in these images (Figure 2a-c and Figure S1) as well as the presence of some structures corresponding to MnP. The EDS shown in Figure 2d confirms the presence of MnP onto the f-MoSe$_2$ flakes. Indeed, carbon, nitrogen and manganese were detected in this kind of regions. Although the ca. 0.28 % atomic Mn is low (see insert of Figure 2d), this is as expected due to the fact that each porphyrin grafted onto $MoSe_2$ contains only one Mn atom.

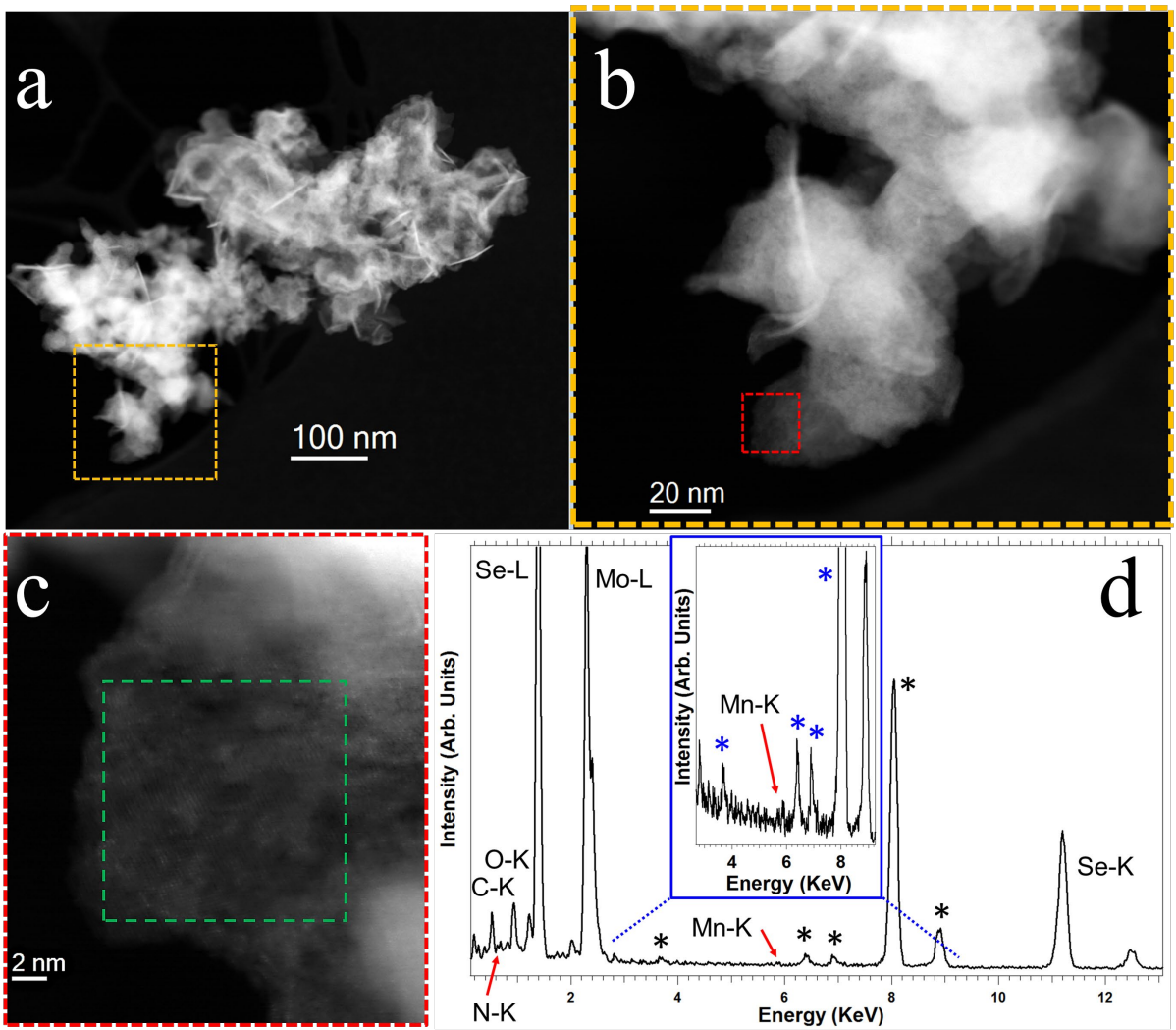

**Figure 2.** (a and b) Low- and medium-magnification high-angle annular dark-field (HAADF)-STEM images recorded on a $MoSe_2$-MnP flake, respectively. (c) High-magnification HAADF-STEM image of the red highlighted area in Figure 2b. (d) EDS spectrum acquired in the green highlighted area of Figure 2c. Signals coming from C, N, Mo, Se and Mn can be observed in this spectrum. Impurities due to Ca, Fe, Co and Cu from the sample holder, objective lens of the microscope and TEM grid are denoted with *. The insert corresponds to a zoom in the EDS spectrum for highlighting the Mn signal.

Steady-state electronic absorption spectroscopy was engaged to document the formation of $MoSe_2$-MnP (Figure 3). Specifically, the UV-Vis spectrum of MnP evidences characteristic absorptions at 343, 374 and 398 nm, at 471 nm, corresponding to the Soret band and at 526, 575 and 613 nm corresponding to the Q bands. In the absorption spectrum of $MoSe_2$-MnP, the Soret band is red-shifted by 6 nm, compared to that due to non-grafted MnP, located at 477 nm. The latter finding states the development of electronic communication between MnP and $MoSe_2$ at the ground state. Furthermore, the Q bands are also evident in the UV-Vis spectrum of $MoSe_2$-MnP at 578 and 612 nm.

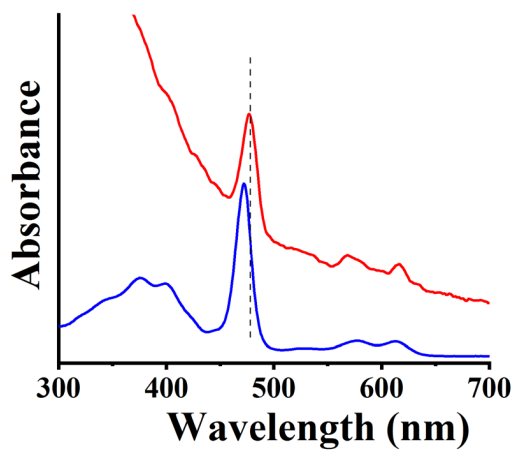

**Figure 3.** UV-Vis spectra for $MoSe_2$-MnP (red) and MnP (blue), in dichloromethane.

The electrocatalytic function of $MoSe_2$-MnP for HER and ORR was investigated. In order to explore the hydrogen evolution reaction from electrochemical water splitting, pseudoelectrodes were constructed by drop-casting dispersions of $MoSe_2$-MnP, along with reference materials and Pt/C benchmark, onto the glassy-carbon electrode. In more detail, linear sweep voltammetry (LSV) assays in aqueous 0.1 M KOH electrolyte (Figure 4a) were performed. Markedly, an onset overpotential at -0.045 V vs RHE was recorded for $MoSe_2$-MnP, which is by 230 mV lower than the one recorded for $MoSe_2$, ca. -0.280 V vs RHE. Additionally, reference materials f-$MoSe_2$ and MnP noted higher onset potential values at -0.5 and -1.2 V, respectively. Thus, incorporation of MnP, via robust covalent anchoring onto MoSe within the $MoSe_2$-MnP hybrid material, resulted in the development of very active electrocatalyst, with catalytic efficiency that exceeds the one of the individual components, highlighting the beneficial role of the metallated porphyrin to enhance the HER electrocatalytic activity of $MoSe_2$.

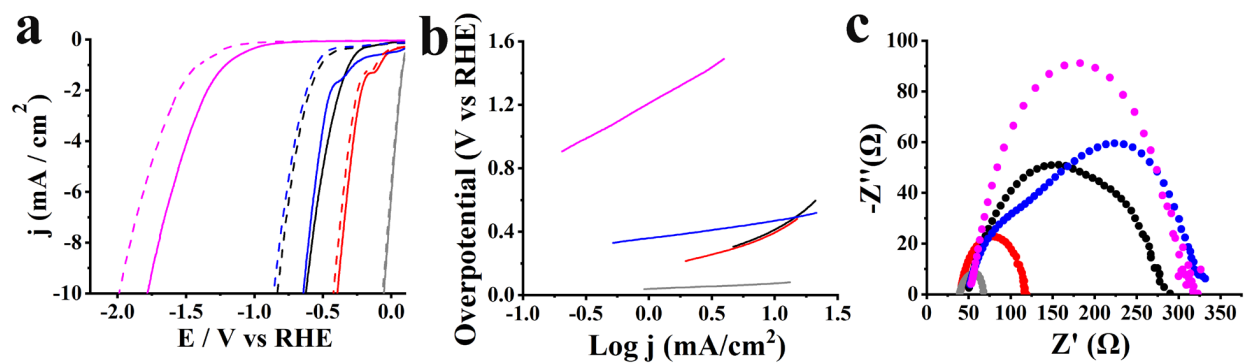

**Figure 4.** (a) LSVs for HER before (solid lines) and after 10,000 cycles (dashed lines), (b) Tafel slopes, and (c) Nyquist plots, for $MoSe_2$-MnP (red), $MoSe_2$ (black), f-$MoSe_2$ (blue), MnP (pink) and Pt/C (grey). The LSVs obtained at 1,600 rpm rotation speed and 5 mV/s scan rate in aqueous 0.1 M KOH.

Electrochemical hydrogen evolution at -10 mA/cm$^2$ is a typical figure of merit to evaluate HER (Figure 4a). Thus, $MoSe_2$-MnP operates the HER at -10 mA/cm$^2$ at -0.39 V vs RHE, ca. 230 mV lower than that of $MoSe_2$, at. -0.61 V vs RHE. For f-$MoSe_2$ and MnP, higher potential values are required to drive HER at the benchmark current density of -10 mA/cm$^2$, at -0.63 and -1.78 V, respectively. The significantly lower overpotential value for driving protons reduction to molecular hydrogen of $MoSe_2$-MnP is related to the presence of the manganese-metallated porphyrin and the synergetic effect with MoSe$_2$, while the covalent conjugation promotes charge-transfer phenomena and current flow.[18]

Furthermore, useful information about the reaction kinetics and mechanism was obtained by extracting Tafel slope values from the LSV polarization curves (Figure 4b). HER in alkaline media is more complex than in acidic electrolyte due to the extra energy that is needed to break the strong covalent H-O-H bond before the adsorption of H*, thus to start the reaction.[35,36] Hence, alkaline HER starts with the adsorption of protons onto the electrocatalyst surface via a reduction process (Volmer step), followed by the evolution of molecular H$_2$, either through the desorption of adsorbed hydrogen atoms onto the electrode (Heyrovsky step) or through the recombination of two adsorbed protons (Tafel step), with an extra dissociation step of water molecules, to produce hydrogen during Volmer and Heyrovsky steps.[35,36] In this manner, Tafel slope value for MoSe$_2$-MnP is 256 mV/dec, indicating that adsorption of protons onto the electrode surface via a reduction process rate limits HER. Meanwhile, Tafel slope values for MoSe$_2$, f-MoSe$_2$ and MnP, ca. 290, 340 and 295 mV/dec, are higher than that of MoSe$_2$-MnP, manifesting the improved HER kinetics within the hybrid.

Additional insight on HER kinetics was brought by electrochemical impedance spectroscopy (EIS) assays. The low frequency region in EIS was employed to determine the charge-transfer resistance ($R_{ct}$) at the interface with the electrolyte, while EIS data were fitted to Randles circuit. In more detail, MoSe$_2$-MnP showed the smallest frequency semicircle in the Nyquist plot, corresponding to a small $R_{ct}$ value of 87 Ω (Figure 4c), compared with the much higher $R_{ct}$ values of 246, 342, and 244 Ω for MoSe$_2$, f-MoSe$_2$ and MnP, respectively. The lower $R_{ct}$ value for MoSe$_2$-MnP reflects the higher conductance and more accessible charge transfer at the electrode/electrolyte interface, due to the covalent linkage of MoSe$_2$ with MnP, favoring the electrocatalytic reaction.

Stability is a key-factor for assessing the performance of electrocatalysts for practical applications. The stability of MoSe$_2$-MnP was assessed, after performing 10,000 ongoing electrocatalytic cycles (Figure 4a). Specifically, MoSe$_2$-MnP proved to be extremely stable after 10,000 cycles by exhibiting a negatively shifted LSV curve of 30 mV (Figure 4a). On the contrary, MoSe$_2$, f-MoSe$_2$ and MnP show higher overpotentials of 220 mV after continues cycling. Table 1 summarizes the electrochemical HER data before and after 10,000 cycles for MoSe$_2$-MnP and compares with those of reference materials and Pt/C. Please note that analysis of the UV-Vis spectrum of MoSe$_2$-MnP after the stability test involving 10,000 cycles did not reveal changes (Figure S2).

**Table 1.** Electrocatalytic HER parameters for MoSe$_2$-MnP in comparison with MoSe$_2$, f-MoSe$_2$, MnP and Pt/C.

| Material | Onset potential (V vs RHE) | Potential (V vs RHE) at -10 mA/cm$^2$ | Tafel slope (mV/dec) | R$_{ct}$ ($\Omega$) |
|---|---|---|---|---|
| MoSe$_2$-MnP | -0.043 | -0.39 | 256 | 87 |
| MoSe$_2$-MnP[a] | -0.068 | -0.42 | 256 | - |
| MoSe$_2$ | -0.28 | -0.62 | 290 | 246 |
| MoSe$_2$[a] | -0.5 | -0.83 | 298 | - |
| f-MoSe$_2$ | -0.4 | -0.63 | 340 | 342 |
| f-MoSe$_2$[a] | -0.55 | -0.85 | 531 | - |
| MnP | -1.2 | -1.78 | 449 | 244 |
| MnP[a] | -1.38 | -1.99 | 485 | - |
| Pt/C | 0.11 | -0.053 | 87 | 33 |
| Pt/C[a] | 0.10 | -0.065 | 87 | - |

[a] after 10,0000 cycles.

In order to scrutinize the 2e$^-$ ORR performance of MoSe$_2$-MnP, LSV measurements on rotating ring disk electrode (RRDE) were employed in O$_2$-saturated aqueous 0.1 M KOH solution at a rotation rate of 1,600 rpm. The electrocatalytic performance for MoSe$_2$-MnP, along with reference materials, is presented in Figure 5a. More precisely, $MoSe_2$-MnP manifests an ORR onset potential at 0.69 V vs RHE, which is 50 mV more positive that $MoSe_2$, f-$MoSe_2$ and MnP, i.e. at 0.64 V vs RHE. In addition, $MoSe_2$-MnP demonstrates more positive half-wave potential value at 0.59 V vs RHE compared to reference materials. More detailed investigations on the $2e^-$ ORR reaction mechanism were carried out by adjusting the rotation rate of RDE. Figure 5b shows the ORR polarization curves at different rotation rates for $MoSe_2$-MnP, revealing that the diffusion-limited current density increases with increasing rotation speed. Additionally, the RRDE approach was applied to shed light in ORR kinetics toward $H_2O_2$ production of $MoSe_2$-MnP (Figure 5c). In more detail, the recorded ring current corresponds to the amount of hydrogen peroxide intermediate produced at the disk electrode during ORR, giving us an accurate way to calculate the number of electrons transferred per oxygen molecule (n) value, but also a reliable method to estimate the percentage (%) of produced $H_2O_2$. In this regard, the electron transfer number for $MoSe_2$-MnP in the potential range from 0.3 to 0.6 V (vs RHE) was determined to be 2.05, manifesting that the oxygen reduction proceeded through the $2e^-$ route, while the high $H_2O_2$ yield of 97% suggests the almost quantitative selective production of hydrogen peroxide. The above results agree with the selectivity calculated based on RDE by K-L plots from the polarization curves at different rotation speeds (Figure S3) as the oxygen reduction proceeded through the $2e^-$ route. Interestingly, $MoSe_2$-MnP presents higher $H_2O_2$ selectivity than other $MoSe_2$-based[38-40] but also $MoS_2$-based ORR electrocatalysts.[41-43]

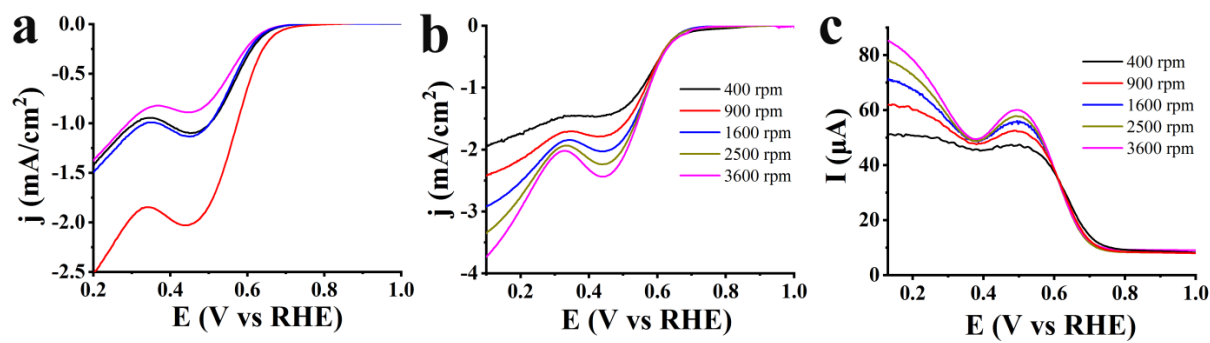

**Figure 5.** (a) ORR polarization curves at 1,600 rpm for MoSe$_2$-MnP (red), MoSe$_2$ (black), f-MoSe$_2$ (blue) and MnP (pink), (b) ORR polarization curves at different rotation rates (400-3,600 rpm) for MoSe$_2$-MnP, and (c) ring response for MoSe$_2$-MnP. All measurements were conducted in O$_2$ saturated aqueous 0.1 M KOH electrolyte and the corresponding LSV polarization curves were recorded at a scan rate of 5 mV/s.

Similarly, the number of electrons transferred was calculated by using both RRDE and RDE for MoSe$_2$ and MnP (Figure S4) and was found to be around 2e$^-$ for both reference materials, while H$_2$O$_2$ yield was determined around 97-98% for both materials (Figure S4). However, MoSe$_2$ and MnP present poor electrocatalytic performance as they show significantly lower onset potential values and current density compared to MoSe$_2$-MnP.

In order to collect additional information on the ORR mechanism, Tafel slopes were extracted from the corresponding LSV curves for all tested materials (Figure 6a). In this context, MoSe$_2$-MnP exhibits slightly lowest Tafel slope value of -52 mV/dec for H$_2$O$_2$ production via ORR compared to f-MoSe$_2$, MnP and MoSe$_2$, being -62, -62 and -60 mv/dec respectively, indicating easier reaction kinetics for MoSe$_2$-MnP. The above values are in agreement with Tafel slope values towards electrocatalytic H$_2$O$_2$ production in alkaline electrolyte found in literature and are consistent with the 2e$^-$ ORR pathway.[37] Furthermore, EIS measurements further confirmed the superior electrocatalytic activity of MoSe$_2$-MnP for H$_2$O$_2$ production. Specifically, MoSe$_2$-MnP showed the smallest $R_{ct}$ value of 732 Ω (Figure 6b) compared with the higher $R_{ct}$ values of 1009, 1010, and 974 Ω for MoSe₂, f-MoSe₂ and MnP, respectively, validating the improved 2e⁻ ORR kinetics within MoSe₂-MnP.

The durability of MoSe₂-MnP was evaluated by performing chronoamperometric assays at a constant applied potential of -0.63 V vs RHE for 10,000 s and rotation speed of 1,600 rpm (Figure 6c). Interestingly, MoSe₂-MnP shows very good stability, with a small current density loss of 13%, possibly due to the formation of MnP aggregates during the durability test.

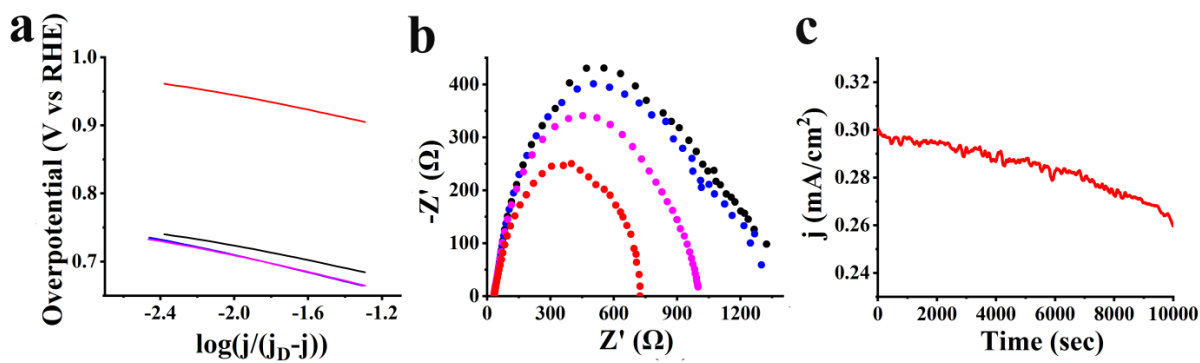

**Figure 6.** (a) Tafel slope and (b) Nyquist plots for MoSe₂-MnP (red), MoSe₂ (black), f-MoSe₂ (blue) and MnP (pink). (c) ORR chronoamperometric response for MoSe₂-MnP at 0.63 V (vs RHE) for 10,000 s.

Finally, in order to get a better understanding of the improved electrocatalytic activity of MoSe₂-MnP, the electrochemically active surface area (ECSA) was estimated. ECSA values were calculated based on the equation[14] ECSA = $C_{dl}/C_{s}$, where $C_{dl}$ stands for the electrochemical double-layer capacitance, while $C_{s}$ is the specific capacitance of a flat surface with 1 cm² of real surface area with a value assumed to be 40 μF/cm² for the flat electrode. For the estimation of the ECSA, cyclic voltamographs were recorded in a non-Faradaic region at scan rates of 50, 100, 200, 300, 400 and 500 mV/sec for all tested materials (Figure S5). ECSA values were obtained from $C_{dl}$ by plotting the $\Delta j = (j_a-j_c)$ at a given potential versus the scan rate as stated in the equation $C_{dl} = d(\Delta j)/2dV_b$. In accordance with the LSV polarization curves for HER and ORR, MoSe$_2$-MnP showed higher ECSA value of ~17.4 cm$^2$. Additionally, MoSe$_2$ and f-MoSe$_2$ present lower but close to each other ECSA, ca. ~12.5 and ~11.18 cm$^2$, respectively. Again, consistent with the electrocatalytic activity for both reactions, MnP shows the lowest ECSA value of ~2 cm$^2$. The above findings are in line with the electrocatalytic results towards HER and ORR, since the higher ECSA values denote efficient accessibility of the electrocatalytic active sites.

**Conclusions**

The combination of MoSe$_2$ with manganese-metallated porphyrin is a great approach to boost their electrocatalytic properties. In this work, we combined the inherent electrocatalytic properties of MoSe$_2$ with the high conductance and electrocatalytic activity of MnP towards the preparation of an inexpensive, highly stable and efficient HER and selective ORR electrocatalyst for H$_2$O$_2$ production. Briefly, MoSe$_2$ was firstly modified with pyridine rings through metal-ligand coordination and afterwards the grafted amine moieties were covalently conjugated with the MnP derivative. This novel modification procedure provides an easy, mild and robust way to functionalize TMDs of both polymorphs. The newly prepared MoSe$_2$-MnP was used to construct pseudoelectrodes and tested as electrocatalyst to reveal that it operates HER 230 mV lower than MoSe$_2$. Additionally, the beneficial role of MnP is highlighted through RRDE studies, showing that ORR follows the 2e$^-$ reduction pathway with high selectivity toward H$_2$O$_2$. Although incorporation of MnP onto MoSe$_2$ enhances the catalytic activity due to the beneficial role of the former, controlling the functionalization of MoSe$_2$ and therefore the amount of MnP incorporated onto MoSe$_2$ for correlating and tuning the catalytic ability of MoSe$_2$-MnP, is not an easy task to perform. All in all, this advanced functionalization approach may open new routes for the realization and exploration of bifunctional electrocatalysts.

ASSOCIATED CONTENT

**Supporting Information**. The Supporting Information is available free of charge on the ACS Publications website. Experimental section including preparation of $MoSe_2$, f-$MoSe_2$, PorMn and $MoSe_2$-MnP, High-magnification HAADF-STEM images, UV-Vis spectra and electrochemical data.

AUTHOR INFORMATION

**Corresponding Author**

E-mail: akagkoura@eie.gr (A. Kagkoura); arenal@unizar.es (Raul Arenal); tagmatar@eie.gr (N. Tagmatarchis)

**Author Contributions**

The manuscript was written through contributions of all authors. All authors have given approval to the final version of the manuscript.

**Acknowledgments**

R.A. acknowledges support from Spanish MICINN (PID2019-104739GB-100/AEI/10.13039/501100011033), Government of Aragon (projects DGA E13-20R) and from EU H2020 "ESTEEM3" (Grant number 823717) and Graphene Flagship (881603). The TEM studies were performed in the Laboratorio de Microscopias Avanzadas (LMA), Universidad de Zaragoza (Spain). R.A. acknowledges support from Spanish MICINN (PID2019-104739GB-

100/AEI/10.13039/501100011033), Government of Aragon (projects DGA E13-20R) and from EU H2020 "ESTEEM3" (Grant number 823717) and Graphene Flagship (881603).